\documentstyle[prl,psfig,aps]{revtex}

\begin{document}
\draft

\narrowtext
\twocolumn

\noindent
{\bf Comment on ``Creating Metastable Schr\"odinger Cat States''}

In a recent Letter~\cite{kn:slomi} Slosser and Milburn have presented
a very interesting model in which homodyne feedback is used to
produce a macroscopic quantum superposition of coherent states inside
an optical cavity. Due to the interaction between the two modes
supported by the cavity, and to the feedback mechanism, such a way
of generating Schr\"odinger cat states turns out to be less
sensitive to decoherence than previously proposed
models~\cite{kn:yusto,kn:miho,kn:dami} which considered the Kerr
effect in a nonlinear medium.
In this Comment I would like to point out some considerations
about the effectiveness of the scheme of Ref.~\cite{kn:slomi}.

\begin{figure}
\centerline{\hbox{\psfig{figure=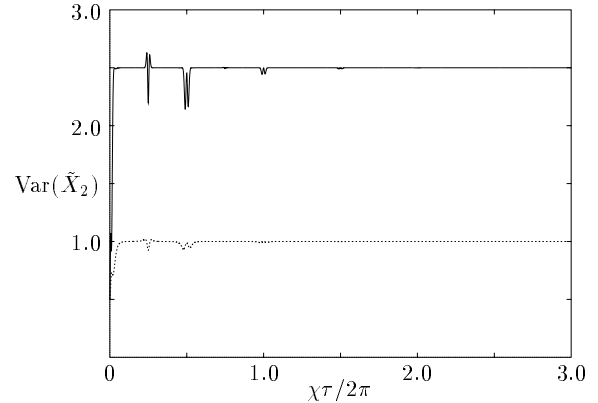,width=3.0in}}}
\vspace{0.2cm}
\caption{Plot of the variance of the approximate quadrature $\tilde{X}_2$
         as a function of time for $\chi=0.3$. Top curve (solid line):
         $|\alpha_0|^2=4.0$; bottom curve (dashed line): $|\alpha_0|^2=1.0$.}
\label{fg:fig}
\end{figure}

The central result of Slosser and Milburn is given by Eq.~(9) of
Ref.~\cite{kn:slomi}, which is a master equation for the mode $a$
alone, after the feedback mechanism and the adiabatic elimination
of mode $b$ have been included. In this equation, $\chi$ is the
feedback gain and can be large. This fact indeed allows the authors
of Ref.~\cite{kn:slomi} to ignore the damping in the dynamics of
mode $a$. This master equation is similar to that for an anharmonic
oscillator coupled to a zero-temperature bath~\cite{kn:miho}, apart
from the phase diffusion term. Such a difference manifests itself,
{\it e.g.}, in the time evolution of the first moment, as illustrated
in Eq.~(13) of Ref.~\cite{kn:slomi}, which we rewrite here for convenience
\begin{equation}
\langle a(\tau) \rangle = \alpha_0\frac{e^{-i\chi\tau}}{[C(\chi\tau)]^2}
e^{-2|\alpha_0|^2[C(\chi\tau)-1]/C(\chi\tau)}e^{-\chi^2\tau}\;,
\label{eq:moment}
\end{equation}
where $|\alpha_0\rangle$ is the initial coherent state and $C(\chi\tau)=
2-\exp(-2i\chi\tau)$.
As a result, the decay rate is independent of the initial photon number,
whereas, for an anharmonic oscillator coupled to a thermal bath, this
would not be the case~\cite{kn:miho,kn:dami}. Moreover, as stated by
Slosser and Milburn, an advantage of the present model should be that
the value of $\chi$ is not any longer related to the nonlinear coupling
of a normal $\chi^{(3)}$ crystal and can be large. However, for large
values of $\chi$, the damping factor $\exp(-\chi^2\tau)$ in
Eq.~(\ref{eq:moment}) causes a fast decay as well.
If one computes the variance
of the approximate quadrature $\tilde{X}_2$, it is easy to recognize
the presence of a similar damping factor. In the plots of Fig.~2 of
\cite{kn:slomi} this is visible, although the authors have chosen
a rather small value $\chi=0.1$. For larger values of $\chi$ the
possibility of the formation of the Yurke-Stoler (YS) state~\cite{kn:yusto}
is washed
out within one recurrence time, similarly to Fig.~2(b) of \cite{kn:slomi}.
This is shown in Fig.~\ref{fg:fig}, where the variance of the approximate
quadrature $\tilde{X}_2$ is plotted versus time for $\chi=0.3$ and
for two values of $|\alpha_0|^2$. Already for this value of $\chi$,
in both cases the generation of a ``quasi'' YS state is possible only
in the first occurrence, for $\chi\tau=\pi/2$: for later times such a
possibility is destroyed by the presence of the damping factor, just
like in the case of an anharmonic oscillator interacting with a
zero-temperature bath, as shown in Fig. 2(b) of \cite{kn:slomi}.
Fig.~\ref{fg:fig} should be contrasted with Fig. 2 of Ref.~\cite{kn:slomi}.
Thus, contrary to Slosser and Milburn's claim, increasing the value of $\chi$
instead of helping the formation of a YS-type state, makes it more and
more difficult.

It seems to turn out that in order to reach the generation of
Schr\"odinger cat states in quantum optics, one has still
to find media for which the
ratio nonlinearity to dissipation is high~\cite{kn:fox,kn:noi}.
Alternatively, one could make use of other
schemes~\cite{kn:atom} which utilize atom-field interactions.

I thank Prof. W. Schleich for the kind hospitality at the University
of Ulm, and acknowledge the European Community (Human Capital and
Mobility programme) for support.

\vspace{0.3cm}

\noindent
Mauro Fortunato \par
Abteilung f\"ur Quantenphysik \par
Universit\"at Ulm, D-89069 Ulm, Germany

\vspace{0.3cm}

\noindent
Received February 13, 1996 \par

\end{document}